\documentclass[aps,pra,reprint,onecolumn,notitlepage,eqsecnum,floatfix,showkeys,nofootinbib]{revtex4-2}

\usepackage{amsmath}
\usepackage{amsbsy}
\usepackage{xcolor}
\usepackage{graphicx}
\usepackage{hyperref}
\usepackage{multirow}
\usepackage{linearb}
\usepackage{relsize}
\usepackage{calrsfs}
\DeclareMathAlphabet\mathbfcal{OMS}{cmsy}{b}{n}

\newcommand{\bq}{\begin{eqnarray}}
\newcommand{\eq}{\end{eqnarray}}
\newcommand{\bqn}{\begin{eqnarray*}}
\newcommand{\eqn}{\end{eqnarray*}}
\newcommand{\bqs}{\begin{subequations}}
\newcommand{\eqs}{\end{subequations}}
\newcommand{\bw}{\begin{widetext}}
\newcommand{\ew}{\end{widetext}}

\newcommand{\aaa}{{\boldsymbol a}}

\newcommand{\BBB}{\mathbfcal{B}}
\newcommand{\kk}{{\boldsymbol k}}

\newcommand{\rr}{{\boldsymbol r}}

\newcommand{\QQ}{{\boldsymbol Q}}
\newcommand{\KK}{{\boldsymbol K}}

\newcommand{\pp}{{\boldsymbol p}}

\newcommand{\nnabla}{{\boldsymbol\nabla}}

\newcommand{\cals}{{\cal S}}
\newcommand{\calz}{{\cal Z}}
\newcommand{\cald}{{\cal D}}

\newcommand{\calh}{{\cal H}}

\begin{document}
\title{Polaron versus Anderson Localization}

\author{Riccardo Fantoni}
\email{riccardo.fantoni@scuola.istruzione.it}
\affiliation{Universit\`a di Trieste, Dipartimento di Fisica, strada
  Costiera 11, 34151 Grignano (Trieste), Italy}

\date{\today}

\begin{abstract}
We compare two kinds of affine localizations in physics: the localization 
in a short range polaron and the one in a Wick rotated Anderson stochastic model. 
The conditions on the interaction potential necessary to see the transnational
symmetry breaking localization phase transition is identical in the two
problems. We therefore suggest that they should belong to the same universality
class of the renormalization group for the localization phase transition. 
\end{abstract}


\maketitle
\section{Introduction}
\label{sec:intro}

In physics, {\sl phase transitions} occur in a many body system only in the 
mathematical thermodynamic limit. In a one body system it is necessary, 
on the other hand, to imagine a {\sl background} that drives the body 
change of behavior in correspondence of the phase transition. In this short 
work we will consider two examples of backgrounds. A deterministic one made 
of a ionic crystal in the thermodynamic limit upon which the polaron moves 
and a stochastic one made of a 
disordered medium with probability amplitudes of sites occupation. The 
former is the problem of a polaron \cite{Frohlich1950} and the latter is 
Wick rotation of the Anderson problem \cite{Anderson1958} where the 
thermodynamic limit is substituted with a probabilistic description of the 
sites occupation. The change 
in behavior for the body observed in both systems is that from an 
{\sl extended state} to a {\sl localized state} where we have a translational 
symmetry breaking. The existence of the phase transition requires a short
range interaction potential in both cases: the retarded potential for the 
polaron and the site-site potential for the Anderson problem.
The transition from one state to the other depends on the temperature 
in both problems and on the coupling constant with the background, in 
the deterministic polaron case, and on the width of the probability 
distribution of the stochastic noise giving rise to the ``kinetic'' 
energy contributing for the jumps between sites, and on the sites density, 
in the Anderson case. In 
particular in order to have localization in the polaron problem the
temperature has to be small and/or the coupling large and in the 
Wick rotated Anderson problem again the temperatures has to be small 
in order to observe the long imaginary time behavior, the variance of the 
noise responsible for the random kick between sites has to be large,
and the sites density has to be large.

We will then describe first the polaron localization problem and then the 
Wick rotated Anderson localization problem and suggest that the two should 
belong to the same universality class of the renormalization group describing 
the physical localization phase transition. The scaling transformation 
generating the semigroup consist in a scaling of the polaron position vector
or of the site probability amplitudes vector. The former is $d$-dimensional 
with $d$ the dimensions of the space in which the polaron moves and the latter
is $N$-dimensional with $N$ the number of sites of the random medium in
the $d$-dimensional space.

It is no surprise that the Anderson problem has played a relevant role in the
modellization of the Metal-to-Insulator-Transition (MIT) 
\cite{Abrahams1979,Cheremisin2017}.

\section{Two kinds of localizations}
\label{sec:localizations}

In this section we compare two kinds of affine localizations in physics: the 
localization in a short range (deterministic) polaron and the one in the 
(Wick rotated) Anderson stochastic model.

\subsection{Polaron localization}
\label{sec:polaron}

A {\sl polaron} is an electron in a ionic crystal of volume $\Omega$. 
The electron polarizes the lattice in its neighborhood. The dispersion in a 
crystal has two branches: an optical branch $\omega(k)=\omega>0$ independent 
of $k$ and an acoustic branch $\omega(k)=vk$, with $v>0$ the sound velocity, 
as $k\to 0$. For concreteness we will carry on our discussion assuming a three 
dimensional crystal.

The Hamiltonian $\calh=\calh_{\rm ele}+\calh_{\rm lat}+\calh_{\rm int}$ for the 
electron of mass $m$ and the lattice is due to Fr\"ohlich 
\cite{Frohlich1950,Frohlich1954,FeynmanFIP}
\bq
\calh=\frac{\hat{\pp}^2}{2m}+
\Omega\int\frac{d\kk}{(2\pi)^3}\hbar\omega(k)a_\kk^\dagger a_\kk +
i\alpha\sqrt{\frac{\hbar}{mv^4}}\Omega\int\frac{d\kk}{(2\pi)^3}\frac{\omega^{3/2}(k)}{k}
\left[a_\kk^\dagger e^{-i\kk\cdot\rr}-a_\kk e^{i\kk\cdot\rr}\right],
\eq
where $\rr$ is the {\sl electron position}, $\hat{\pp}=-i\hbar\nnabla$ its conjugate 
momentum, $a_\kk^\dagger, a_\kk$ are the creation and annihilation operators for 
a {\sl phonon} of dispersion relation $\omega(k)$ and momentum $\hbar\kk$, and 
$\alpha$ is the adimensional coupling constant. 
We will adopt units such as $\hbar=m=v=1$.

Next we recall that the positions and momenta of the phonons are given by
\bq
q_\kk &=&\sqrt{\frac{1}{2\omega(k)}}(a_\kk^\dagger+a_{-\kk}),\\
p_\kk &=&i\sqrt{\frac{\omega(k)}{2}}(a_{-\kk}^\dagger-a_{\kk}),
\eq
defining $a'_\kk=-ia_{-\kk}$ we find $q'_\kk=p_\kk$ and $p'_\kk=-q_\kk$ above and
we can rewrite, dropping the primes,
\bq
\calh=\frac{\hat{\pp}^2}{2}+\Omega\int\frac{d\kk}{(2\pi)^3}\frac{1}{2}
\left[p_\kk^2+\omega(k)^2q_\kk^2\right]+\sqrt{2}\alpha\Omega\int\frac{d\kk}{(2\pi)^3}
\frac{\omega^2(k)}{k}q_\kk e^{i\kk\cdot\rr}, 
\eq
where we accomplished the task of rewriting the interaction term as a function
of the electron and phonons positions only.

Assume now that the system is in thermal equilibrium at an inverse temperature
$\beta=1/k_B T$ with $k_B$ the Boltzmann constant and $T$ the absolute temperature. 
We will also assume to be at very low temperature.
We can then use a path integral \cite{FeynmanFIP} to manipulate the polaron 
density matrix $\rho=e^{-\beta\calh}$ and write the partition function
\bq \label{eq:Z}
\calz={\rm Tr}\left(e^{-\beta\calh}\right)=\int_{\substack{\rr(0)=\rr(\beta)\\q_i(0)=q_i(\beta)}}
e^{-\cals}\,\cald \rr(u)\cald q_1(u)\cald q_2(u)\cdots,
\eq
where ${\rm Tr}(\cdot)$ is the operator trace and the action integral $\cals$ is
\bq \label{eq:action}
\cals=\int_0^\beta\left\{\frac{\dot{\rr}^2(u)}{2}+
\Omega\int\frac{d\kk}{(2\pi)^3}\frac{1}{2}
\left[\dot{q}_\kk^2+\omega(k)^2q_\kk^2\right]+\sqrt{2}\alpha\Omega\int\frac{d\kk}{(2\pi)^3}
\frac{\omega^2(k)}{k}q_\kk e^{i\kk\cdot\rr}\right\}\,du.
\eq
The path integral over the phonons in (\ref{eq:Z}) can easily be performed
\cite{FeynmanFIP} because $\dot{q}_\kk$ and $q_\kk$ both appear quadratically
and linearly in the action (\ref{eq:action}). The result is
\footnote{Here we are assuming to be working at low temperature when $\beta$ is large. 
The exact result would require to substitute the term $e^{-\omega(k)|t-s|}$ with
$e^{-\omega(k)|t-s|}/[1-e^{-\omega(k)\beta}]+e^{\omega(k)|t-s|}e^{-\omega(k)\beta}/[1-e^{-\omega(k)\beta}]$.}
\bq
\calz&=&{\rm Tr}\left(e^{-\beta\calh}\right)=\int_{\rr(0)=\rr(\beta)}
e^{-S_H}\,\cald \rr(u),\\
S_H&=&\frac{1}{2}\int_0^\beta\dot{\rr}^2(u)\,du-\frac{\alpha^2}{2}\int_0^\beta\int_0^\beta
\Omega\int\frac{d\kk}{(2\pi)^3}\frac{\omega^3(k)}{k^2}e^{i\kk\cdot[\rr(t)-\rr(s)]}
e^{-\omega(k)|t-s|}\,dt ds.
\eq 

For example for an optical polaron one finds \cite{FeynmanFIP} for the effective 
{\sl retarded} interaction potential,
\bq
V^{\rm eff}_{\rm opt}=-\frac{\Omega\alpha^2\omega^3}{8\pi}
\frac{e^{-\omega|t-s|}}{|\rr(s)-\rr(t)|},
\eq
whereas for an acoustic polaron one would get \cite{Fantoni12d,Fantoni13a},
\bq
V^{\rm eff}_{\rm aco}=-\frac{\Omega\alpha^2}{(2\pi)^2}\frac{1}{|\rr(s)-\rr(t)|}
\int_0^{k_0}dk\,k^2\sin(k|\rr(s)-\rr(t)|)e^{-k|t-s|},
\eq
where $k_0$ is the Debye cutoff. 

But one is free to choose even more exotic dispersion relations. For example for
$\omega^3(k)/k^2\propto k^{\gamma-d}$ as $k\to 0$ then 
$V^{\rm eff}\propto \Delta r^{-\gamma}$ for $\Delta r=|\rr(s)-\rr(t)|\to\infty$
with $0<\gamma<d, \gamma\neq d-2$ where $d$ is the space dimension. 
Moreover $V^{\rm eff}$ will decay faster than any inverse power of $\Delta r$
whenever $\omega^3(k)/k^2$ is analytic as $k\to 0$ \cite{Lighthill}.

The effective Hamiltonian for the polaron, after tracing out the phonons degrees 
of freedom, would then be
\bq
H^{\rm pol}=\frac{\hat{p}^2}{2}+V^{\rm eff}(\Delta r,\Delta t).
\eq
with $\Delta r=|\rr-\rr'|$ and $\Delta t=|t-t'|$. And the density matrix satisfies 
the effective Bloch equation
\bq \label{eq:b-polaron}
-\frac{\partial\rho}{\partial\beta}&=&H^{\rm pol}\rho.
\eq
where the coordinate representation of the density matrix $\rho(\rr,\rr';\beta)$ 
satisfies the initial condition $\rho(\rr,\rr';0)=\delta(\rr-\rr')$.

\subsection{Anderson localization}
\label{sec:Andreson}

We may think at the effective retarted interaction potential of the 
polaron problem as the Anderson site-site potential $V^{\rm And}$ 
\cite{Anderson1958} which determines the dynamics of his {\sl probability 
amplitude} $a(\rr,t)$. On the continuum, Anderson equation reads
\bq \label{eq:Anderson}
i\frac{\partial a(\rr,t)}{\partial t}=H^{\rm And}a(\rr,t)=
E(\rr,t)a(\rr,t)+\int d\rr'\, V^{\rm And}(|\rr-\rr'|)a(\rr',t),
\eq
where $E(\rr,t)$ is a stochastic variable
\footnote{Note that Anderson chooses $E$ independent of time \cite{Anderson1958}
whereas we here adopt this more general point of view. So in Anderson point
of view we would not need to specify a conditional probability any further.}
with unconditional probability $p(E,\rr;t)dE$ that $E(\rr,t)$ assumes values 
in $[E(\rr,t),E(\rr,t)+dE]$. This probability distribution has a width 
$W$. Anderson discretizes the $d$-dimensional space 
$\rr=(x_1,x_2,\ldots,x_d)$ into a lattice made of $N$ sites at $\rr_j$ for
$j=1,2,\ldots,N$ in a volume $\Omega$, with sites density $n=N/\Omega$.

When studied in imaginary time $t\to -i\beta$, with $\beta=1/k_BT$ the inverse 
temperature, through a Wick rotation, 
Anderson equation (\ref{eq:Anderson}) can be thought as the Bloch equation 
for a thermal density matrix $a_j$ at each lattice site. 
Also, it is a Langevin equation where the first term on the right hand side is 
a noise term and the second term is a drift term
\bq \label{eq:sde}
-\frac{\partial a_j(\beta)}{\partial\beta}=E_j(\beta)a_j(\beta)+
\sum_{i\neq j}V^{\rm And}_{ji}a_i(\beta).
\eq

We can then use Ito calculus \cite{Gardiner} to write the Fokker-Planck 
equation corresponding to the stochastic differential equation (\ref{eq:sde}), 
which describes the time evolution of the transition 
probability $p(\aaa,\aaa^0;\beta-\beta^0)$ that the stochastic process assumes 
the values $\aaa=(a_1,a_2,\ldots,a_N)$ at time $\beta$ when it had assumed the 
values $\aaa^0$ at time $\beta^0$. 

Assuming that the noise term is characterized by Gaussian white noise 
$E_j=\xi_j$ and calling $W=\zeta$ (see Appendix \ref{app:wn}) one finds \cite{Gardiner}
\footnote{Note, that the Fokker-Planck equation (\ref{eq:FP}) holds for the 
stochastic differential equation (\ref{eq:sde}) only within the framework 
of Ito calculus. The relation between (\ref{eq:sde}) and the Fokker-Planck
equation is slightly different when Stratonovitch calculus is applied. Note 
also that the amplitude $\zeta$ of the stochastic kick does not enter in 
(\ref{eq:FP}) for our particular choice of the Gaussian white noise, but 
for more general stochastic variables $E_j(\beta)$ this may not be 
the case anymore.}
\bq \label{eq:FP}
\frac{\partial p(\aaa,\aaa^0;\beta-\beta^0)}{\partial\beta}&=&
\Lambda p(\aaa,\aaa^0;\beta-\beta^0),\\ \label{eq:L}
\Lambda &=& \sum_{i\neq j}V_{ji}^{\rm And}a_i\frac{\partial}{\partial a_j}
+\sum_i\frac{\partial^2}{\partial a_i^2}a_i^2,\\
p(\aaa^0,\aaa^0;0)&=&1,
\eq
Where $\Lambda$ is made up of two differential operators acting on $p$. 
The unconditional probability for the realization
of $\aaa^1,\aaa^2,\ldots$ at times $\beta^1,\beta^2,\ldots$ will then be
$p(\aaa^0,\aaa^1,\aaa^2,\ldots)=p(\aaa^0;\beta^0)p(\aaa^1,\aaa^0;\beta^1-\beta^0)
p(\aaa^2,\aaa^1;\beta^2-\beta^1)\ldots$ with $p(\aaa^0;\beta^0)$ the unconditional 
probability at the initial time. Given the unconditional probability and 
the transition probability we can completely characterize the Markov 
process. In particular we will be interested in the long imaginary time,
or low temperature, regime.

In Ref. \cite{Fantoni12d,Fantoni13a} we studied the low temperature 
properties of an acoustic polaron through Path Integral Monte Carlo (PIMC)
and we found the existence of a phase transition from an extended state 
to a localized state as the phonons-electron coupling constant $\alpha$ 
is increased at constant temperature or as the temperature is decreased 
at constant $\alpha$.

Comparing the polaron and the Wick rotated Anderson problems we see that the 
interaction terms in the stochastic differential equation (\ref{eq:sde}) 
and in the partial differential Bloch equation (\ref{eq:b-polaron}) are similar 
and the kinetic term is deterministic in the polaron case and stochastic 
in the Anderson case. Therefore the probability amplitude in the Anderson 
problem, $\aaa$, is a stochastic vector determined by the conditional probability
$p(\aaa,\aaa';\beta)$. This probability distribution in the Anderson problem 
and the coordinate representation of the density matrix in 
the polaron problem, $\rho(\rr,\rr';\beta)$, play parallel roles. 
We may then propose the following identification between the two problems
\bq \label{eq:id}
p\leftrightarrow\rho~~~~~~\Lambda\leftrightarrow H^{\rm pol}~~~~~~\aaa\leftrightarrow\rr,
\eq
where the probability amplitude vector $\aaa$ in the Anderson problem 
plays the role of the polaron position vector $\rr$. In the two problems,
the solution 
of the Bloch equation for the density matrix of the polaron problem and the
solution of the Fokker-Planck equation for the conditional probability
of the amplitude vector $\aaa$ require a path integral. In fact if one is 
only interested in the probability distribution of the ending value of the 
probability amplitude vector at a large $\beta$ one has to compute
\bq \label{eq:pisda1}
p(\aaa;\beta)&=&\int d\aaa^0d\aaa^1d\aaa^2\ldots p(\aaa^0;\beta^0)
p(\aaa^1,\aaa^0;\Delta\beta)p(\aaa^2,\aaa^1;\Delta\beta)\ldots 
p(\aaa,\aaa^{M-1};\Delta\beta)\\ \label{eq:pisda2}
&=&\int d\aaa^0 p(\aaa^0;\beta^0)
\int_{\substack{\aaa(\beta^0)=\aaa^0\\\aaa(\beta)=\aaa}} 
e^{-S_\Lambda}\cald \aaa(t),
\eq
where in Eq. (\ref{eq:pisda1}) we discretized also the imaginary time so that 
$\Delta\beta=(\beta-\beta^0)/M=\tau$. In the limit of $M$ very large we can 
rewrite this equation in terms of a path integral as in Eq. (\ref{eq:pisda2}) 
where we indicated with $S_\Lambda$ the ``action'' which determines the dynamics 
(\ref{eq:FP}) of the stochastic probability amplitude vector $\aaa$
\footnote{Note that the use of Trotter formula \cite{Cohen1982} requires that 
the two addend operators in $\Lambda$ of Eq. (\ref{eq:L}) be bounded from below.}.
As is shown in Appendix \ref{app:SL} the short time propagator 
$p(\aaa,\aaa';\tau)$ is made up of three terms, two delta functions and a 
Gaussian. In reconstructing the path integral, the product of all the 
first delta functions in Eq. (\ref{eq:papp}) will simply contribute to 
the probability for the path to remain fixed at $\aaa_0$. The weight of 
the product of these deltas contains powers of $e^N$ which suggests that 
a high $N$ at fixed volume $\Omega$ favors localization and no diffusion 
can take place
\footnote{Note that in the real time Anderson problem the exact opposite 
behavior takes place. It is at small sites density $n=N/\Omega$ that no 
diffusion takes place.}. 
All the other product terms will be responsible for the probability 
of a jump between one site and another. The second delta 
in Equation (\ref{eq:papp}) contains the site-site interaction 
potential alone and the third term is a purely diffusion one.
So we see how the Anderson site-site interaction potential is 
responsible for reshuffling of the site probability amplitudes
\footnote{If $\bar{V}^{\rm And}_{ij}$ is the matrix $V^{\rm And}_{ij}$ 
with all $4+1/\tau$ on the diagonal then if 
${\rm det}\{\bar{V}^{\rm And}\}\neq 0$ there will be a one to one 
correspondence between $\aaa$ and $\aaa'$.}.

There is no exact analytic solution for the polaron
problem, so one must resort to an exact numerical solution like 
PIMC for example. In Ref. \cite{Fantoni12d,Fantoni13a} we performed
some PIMC simulations showing that an acoustic polaron path $\rr(t)$ 
becomes localized as one lowers the temperature, $\beta$ large, or
increases the coupling constant, $\alpha$ large, as predicted in 1933 
by Landau \cite{Landau1933,Pekar1946,Landau1948}. The localization has 
not yet been observed in an optical
polaron. It is believed that the existence of a localized or self-trapped
path depends crucially on the dispersion relation. In the optical case 
$\omega^3(k)/k^2\propto 1/k^2$ which is the special Coulombic case
among the various long range cases which seem to prevent the 
self-trapped-state (TS) in favor of an extended-state (ES). On the other 
hand it is believed that a dispersion relation giving rise to a short 
range effective interaction, 
$V^{\rm eff}\propto\Delta r^{-\gamma}$ with $\gamma>d$,
would sustain the localization. This same connection that occurs between 
the existence of the translational symmetry breaking between the ES and the 
TS and the asymptotic behavior of the effective interaction in the 
polaron problem {\sl surprisingly} holds also in the real time Anderson 
problem \citep{Anderson1958} and so also in its Wick rotation in 
imaginary time $\beta$. Here the path $\aaa(\beta)$ is described by stochastic 
jumps amongst the different lattice sites and the localization occurs
when $\sum_{j=1}^Na_j(\beta)j$ remains bounded from above, uniformly in
$\beta$.

Recent work has shown that a non interacting Anderson localized system 
can become many body localized. On the other hand theoretical treatments 
have been extended from one polaron to many polaron systems 
\cite{Devreese2005,Bassani2003,Hohenadler2007}. Already two polarons
are expected to have counter intuitive properties like attraction between
the two electrons due to the energy lowering as the two electrons come
together therefore sharing their deformations of the underlying crystal.
This could lead to a bound bipolaron. For strong attraction, bipolarons 
may be small. Small bipolarons have integer spin and thus share some of 
the properties of bosons. If many bipolarons form without coming too 
close, they might be able to form a Bose-Einstein condensate. This has 
led to a suggestion that bipolarons could be a possible mechanism for 
high-temperature superconductivity.

\section{Renormalization group}
\label{sec:rg}

The findings of the previous section should not be so surprising after all 
and certainly suggest that the polaron localization problem and the Anderson 
localization problem belong to the same {\sl universality
class} of the Renormalization Group (RG) \cite{Fisher-RMP1998} with a {\sl 
fixed point} corresponding to the localization phase transition. We can 
then think of a {\sl reduced Hamiltonians} very large space $\calh[\QQ;\KK]$ to
which the two {\sl physical Hamiltonians} $H^{\rm pol}$ and $\Lambda$ belong. Where
$\QQ$ stands for the ``coordinate variables'', $\rr$ for the polaron problem 
and $\aaa$ for the Anderson problem, and $\KK$ are the various 
{\sl thermodynamic fields} like the temperature that can be controlled 
directly by the experimenter and others that embody details of the physical
system that are {\sl fixed by nature}, like the coupling constant $\alpha$ for 
the polaron or the density $n$ and the noise width $W$ for the Anderson problem. 
At the heart of RG theory there is the 
renormalization of the ``coordinates'' via $\QQ\to\QQ'=b\QQ$ with $b>1$
\footnote{Note that for the polaron localization problem iteration of 
such scaling would simply bring the TS into an ES and in the Anderson 
localization problem it will affect the site probability amplitudes again 
producing delocalization.}
. Iterating this scaling transformation $n$ times we can introduce the 
{\sl flow parameter} $n=\log_b(Q'/Q)\geq 0$ from which we can derive a 
{\sl differential} or continuous RG flow as $d\calh/dn=\BBB\calh$. 
So that starting from the physical critical point on the {\sl physical 
manifold} $\calh$, after the $n$th iteration we reach the $n$th 
renormalized critical point for $\calh^{(n)}[\QQ^{(n)};\KK^{(n)}]$ on the 
$n$th renormalized manifold $\calh^{(n)}$. Repeating many times the scaling 
transformation we can then follow the critical trajectories of the various 
subsequent renormalized critical points and eventually reach a {\sl nontrivial} 
fixed point $\calh^*$ such that $\BBB\calh^*=0$. All physical Hamiltonians whose
critical trajectories converge to the same fixed point belong to the same
universality class. We can then choose an expansion 
\cite{Wilson1971a,Wilson1974,Fisher1974b,Wegner1972a,Wegner1972b,Wegner1976,Kadanoff1976}
on the {\sl tangent space} 
to the smooth Hamiltonian space at the fixed point $\calh^*$ to determine the 
various {\sl critical exponents}. In fact we expect that something similar 
to what happens for the (short range) Ising model, which share the same 
critical exponents \cite{Domb1960}, should also happen here with the usual 
strong dependence on dimensionality 
\footnote{For example 2 is expected to be the 
lower critical dimension of the real time Anderson localization problem.}. 
One could, for example, use the RG scaling transformation to calculate the
negative maximal Lyapunov exponent \cite{Lyapunov} attracting the RG critical 
flow to the fixed point \cite{Collet1989}.

\section{Conclusions}
\label{sec:conclusions}

We compared the localization that occurs in a short range polaron and 
the Anderson localization that occurs for a short range site-site interaction.
We put the two systems side by side on the ``mirror'' of the renormalization 
group suggesting that they should belong to the same universality class. 
It would be extremely interesting to carry out a PIMC calculation for the 
Wick rotated Anderson problem so to assess unambiguously its localization
properties. From the other side it would be interesting to find other 
physical Hamiltonians whose critical flows for the translational symmetry 
breaking of the localization phase transition fall on the same fixed point. 
It would also be interesting to study the dependence on dimensionality of 
the fixed point in all these problems and calculate their negative maximal
Lyapunov exponent.  

\acknowledgments
I would like to dedicate this little work to three of my professors. In order,
David M. Ceperley for teaching me the path integral Monte Carlo method, Klaus
J. Schulten for teaching me the stochastic differential equations, and Myron B. 
Salamon for carrying out some enlightening experiments on localization. During
my stay in Urbana from 1995 to 2000.

\appendix
\section{Gaussian noise}
\label{app:wn}

An important idealized stochastic process is the so-called Gaussian white 
noise. This process, denoted by $\xi_i$, is not characterized through 
unconditional and conditional probabilities, but only through the following 
statistical moment and correlation 
function
\bq \label{eq:wn1}
\langle\xi_i(\beta)\rangle &=& 0,\\ \label{eq:wn2}
\langle\xi_i(\beta)\xi_j(\beta')\rangle &=& \zeta^2\delta(\beta-\beta')\,\delta_{ij},
\eq
the term {\sl Gaussian} implies that all cumulants higher than of second order 
are 0. The term {\sl white} is connected with the fact that the imaginary time 
Fourier transform of (\ref{eq:wn2}) is a constant, i.e., entails all frequencies 
with equal amplitude just as white radiation. The importance of the process $\xi$ 
stems from the fact that many other stochastic processes are described through 
stochastic differential equations with a (white) noise term $\xi$. The integral 
of white noise over imaginary time is a Wiener process 
$\omega(\beta)$ with $d\omega=\xi d\beta$ and unconditional and conditional 
probabilities distributions
\bq
p_{\rm W}(\omega;\beta)&=&\frac{1}{\sqrt{2\pi\zeta^2\beta}}e^{-\omega^2/2\zeta^2\beta},\\
p_{\rm W}(\omega,\omega';\beta-\beta')&=&\frac{1}{\sqrt{2\pi\zeta^2\Delta\beta}}e^{-(\Delta\omega)^2/2\zeta^2\Delta\beta},
\eq
with $\Delta\omega=\omega-\omega'$ and $\Delta\beta=\beta-\beta'>0$. So that
\bq \label{eq:W1}
\langle\omega_i\rangle &=& 0,\\ \label{eq:W2}
\langle\omega_i(\beta)\omega_j(\beta')\rangle &=& \zeta^2\beta'\,\delta_{ij},
\eq
and again all cumulants higher than of second order are 0. It is this last 
process that enters into Ito calculus \cite{Gardiner} integrating 
in imaginary time Anderson stochastic differential equation (\ref{eq:sde}).

\section{Anderson ``action''}
\label{app:SL}

Using, as usual, the set of orthonormal plane waves, we find for the Anderson 
short time conditional probability in Eq. (\ref{eq:pisda1}) the following 
expression
\bq \label{eq:papp}
p(\aaa,\aaa';\tau)\propto \prod_{i=1}^N \left\{e^{\tau 2N}\delta(a_i-a_i')+
\delta\left[(1+4\tau)a_i+\tau\sum_{i\neq j}V_{ij}^{\rm And}a_j-a_i'\right]+
\sqrt{\frac{2\pi}{\tau a_i^2}}e^{-(1-a_i'/a_i)^2/\tau}\right\},
\eq
where $\delta$ is a one dimensional Dirac delta. 

The unconditional probability of Eq. (\ref{eq:pisda2}) and the averages
stemming from it can then be numerically computed exactly through the 
Path Integral Monte Carlo (PIMC) method \cite{Ceperley1995}, for example.

\section*{Author declarations}

\subsection*{Conflicts of interest}
None declared.

\subsection*{Data availability}
The data that support the findings of this study are available from the 
corresponding author upon reasonable request.

\subsection*{Funding}
None declared.

\bibliography{loc}

\end{document}